\documentclass[a4paper]{jpconf}
\usepackage{graphicx}
\usepackage{amssymb}
\begin{document}
\def\diffflux{\mbox{GeV}^{-1}\,\mbox{cm}^{-2}\,\mbox{s}^{-1}\,\mbox{sr}^{-1}}
\def\pointflux{\mbox{GeV}^{-1}\,\mbox{cm}^{-2}\,\mbox{s}^{-1}}
\def\diffunits{\mbox{GeV cm}^{-2}\,\mbox{s}^{-1}\,\mbox{sr}^{-1}}
\def\pointunits{\mbox{GeV cm}^{-2}\,\mbox{s}^{-1}}
\def\en{E_{\nu}}
\def\eg{E_{\gamma}}
\def\ep{E_{p}}
\title{Status of neutrino astronomy}

\author{Julia K.~Becker}

\address{Physics Institute, Gothenburg University, 41296 Gothenburg,
  Sweden\\ Physics Department, University of
  Technology Dortmund, 44221 Dortmund, Germany}

\ead{julia.becker@physics.gu.se}
\begin{abstract}
Astrophysical neutrinos can be produced in proton interactions of charged cosmic rays with
ambient photon or baryonic fields. Cosmic rays are observed in balloon, satellite and air 
shower experiments every day, from below $10^{9}$~eV up to macroscopic
energies of $10^{21}$~eV. The observation of different photon fields
has been
done ever since, today with detections ranging from radio wavelengths up to very
high-energy photons in the TeV range. The leading question for neutrino astronomers is now which sources 
provide a combination of efficient proton acceleration with sufficiently high
photon fields or baryonic targets at the same time in order to produce a neutrino flux that is high 
enough to exceed the background of atmospheric neutrinos. There
are only two confirmed astrophysical neutrino sources up to today: the sun and 
SuperNova 1987A emit and emitted neutrinos at MeV energies. 
The aim of large underground Cherenkov telescopes like IceCube and KM3NeT 
is the detection of neutrinos at energies above 100 GeV. In this paper, recent 
developments of neutrino flux modeling for the most promising extragalactic 
sources, gamma ray bursts and active galactic nuclei, are presented.
\end{abstract}
\section{Introduction}
\begin{figure}
\centering{
\includegraphics[width=7.5cm]{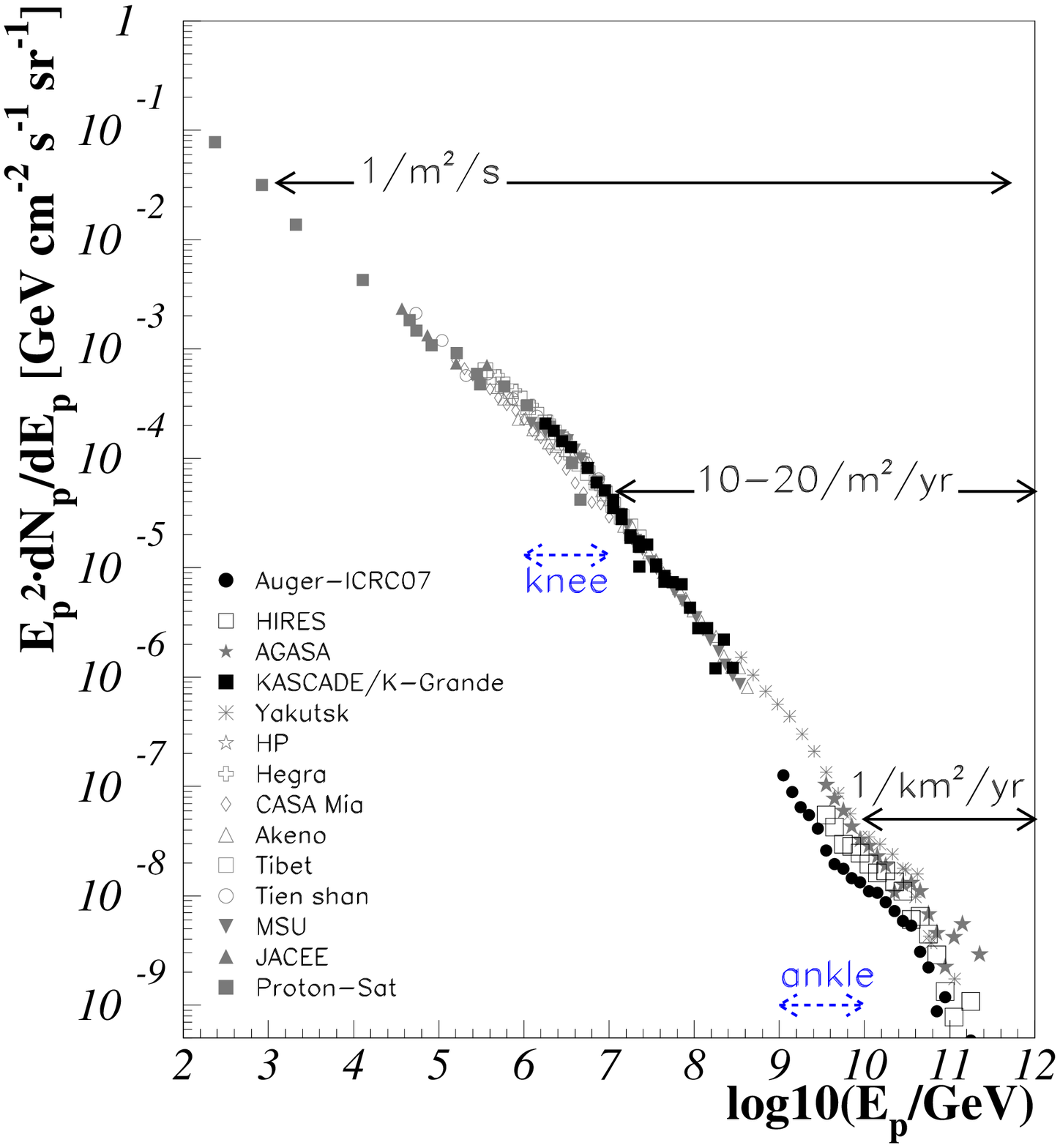}
\includegraphics[width=7.5cm]{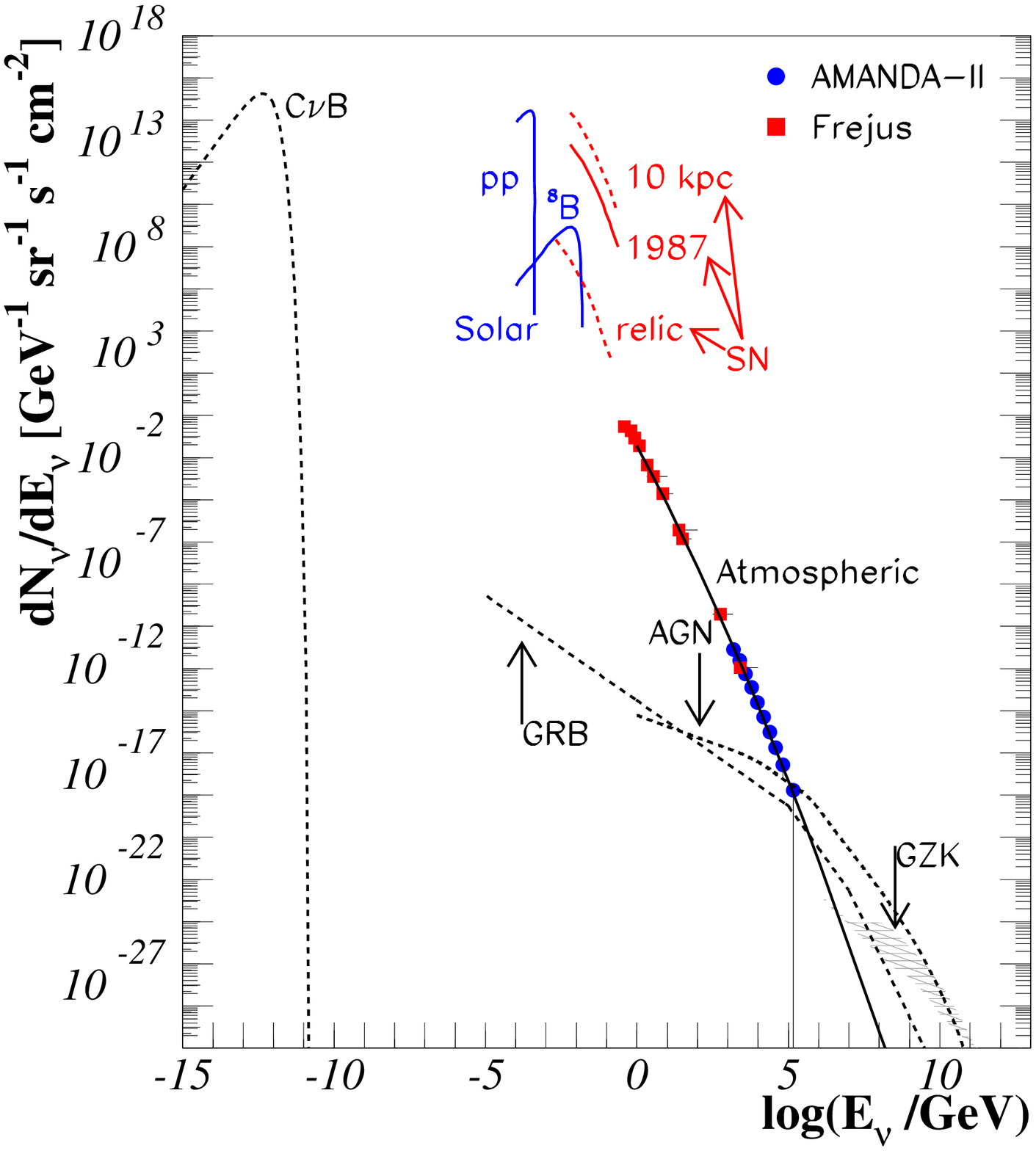}
\caption{The measured spectrum of cosmic rays (left) and the
  predicted spectra of astrophysical neutrinos (right). Cosmic rays
  have been measured since the early 20th century, but their sources
  could not be identified unambiguously yet. Astrophysical neutrinos,
  on the other hand, are predicted from a wide range of sources, but
  difficult to detect. Confirmed sources (solid lines) are the sun
  and the supernova 1987A. See \cite{julias_review}
  and references therein.\label{crs_and_nus:fig}}
}
\end{figure}

The detection of cosmic radiation from space in the early 20th century
was the first step of the today
established idea that particles can be accelerated to ultra-high
energies by cosmic accelerators~\cite{hess1912}. Today, the cosmic ray
energy spectrum ranges from below $10^{9}$~eV up to energies larger
than a joule, $10^{21}$~eV, see Fig.~\ref{crs_and_nus:fig} (left). Still, the origin
of those cosmic rays could not be identified unambiguously
yet due to the scrambling of their original directions by galactic magnetic fields. High-energy neutrinos play an important role in identifying
cosmic ray accelerators. Undeflected and unabsorbed, they traverse the
universe and are therefore perfect messengers from the acceleration
region of baryonic particles. This advantage poses a disadvantage for
the detection of neutrinos, which is one of the reasons why only two neutrino
source predictions could be confirmed until today: solar neutrinos and
the supernova 1987A. Figure~\ref{crs_and_nus:fig} shows the expected
astrophysical neutrino spectrum, ranging from $10^{-3}$~eV energies (cosmic
background neutrinos, decoupled after $1$~s after the Big Bang), up to
$10^{19}$~eV - those high-energy neutrinos connected to the highest
energy cosmic rays. In this contribution, the possible production
sites in astrophysical source classes are the topic of this
paper. Section~\ref{nu_emitters:sec} discusses candidate sources for
high-energy production. In Section~\ref{nu_models:sec}, 
neutrino flux models are presented in the context of current
experimental limits. Section~\ref{outlook:sec} gives an outlook on
future neutrino astronomy.
\newpage
\section{Potential neutrino emitters\label{nu_emitters:sec}}
Neutrinos are dominantly produced in interactions of protons with ambient
photon or proton fields: $p\, \gamma \longrightarrow
\Delta^{+}\rightarrow p\,\pi^{0}/n\,  \pi^{+}$ and $p\, p \rightarrow\pi^{+}\,\pi^{-}\,\pi^{0}$.
The charged pions subsequently decay to produce neutrinos, $\pi^{+}\rightarrow\mu^{+}\,\nu_{\mu}\rightarrow
e^{+}\,\nu_{e}\,\overline{\nu}_{\mu}\,\nu_{\mu}$ and $\pi^{-}\rightarrow\mu^{-}\,\overline{\nu}_{\mu}\rightarrow e^{-}\,\overline{\nu}_{e}\,\nu_{\mu}\,\overline{\nu}_{\mu}$.
The neutrino flavors are
produced at the source in ratios $(\nu_{e}:\nu_{\mu}:\nu_{\tau})=(\overline{\nu}_{e}:\overline{\nu}_{\mu}:\overline{\nu}_{\tau})=(1:2:0)$.
Due to neutrino oscillations, the ratio at Earth becomes $(\nu_{e}:\nu_{\mu}:\nu_{\tau})=(\overline{\nu}_{e}:\overline{\nu}_{\mu}:\overline{\nu}_{\tau})=(1:1:1)$.
In order to produce a significant neutrino flux, two criteria need to be
fulfilled:
\begin{enumerate}
\item The kinematics must allow for the production of the Delta-resonance in
  the case of $p\,\gamma$ interactions, and for the production of the three
  pions in case of proton-proton interactions.
\item The environment of the
source needs to provide sufficient photon or proton targets, i.e.\ the optical
depth for proton-proton or proton-photon interactions, $\tau_{p\,p}$
or $\tau_{p\,\gamma}$, needs to be close to or above unity,
$\tau_{p\,i}=L/\lambda_{mfp}=l\cdot n_{i}\cdot \sigma_{p\,i}\,,$
with $i=p,\,\gamma$. The mean free path is given as
$\lambda_{mfp}=(n_{i}\cdot \sigma_{p\,i})^{-1}$ and $L$ is the extension of the
source. 
\end{enumerate}
In any case, the sources need to contain high-energy protons in order to have
the above collisions. This means the proposed sources for cosmic ray
acceleration corresponds to the predicted sources of neutrino emission. Here,
two criteria have to be met to have a source class responsible for a particle
flux as high as observed in the cosmic ray spectrum:
\begin{enumerate}
\item The maximum energy of the particle must be high enough. A single
  particle's energy is mainly determined by the product of the extension of
  the source, $L$, and the source's magnetic field, $B$, see~\cite{hillas1984}:
$E_{\max}\approx 10^{18}\,{\rm eV}\cdot Z\cdot \beta_{SH}\cdot L/(1\,{\rm
    kpc})\cdot B/(1\,\mu{\rm G})$.
Here, $Z$ is the charge of the particle in units of the electron's charge. Further,
$\beta_{SH}$ is the velocity of the shock region, in which the particle is
accelerated, in units of the speed of light.
\item The non-thermal, electromagnetic luminosity of the sources is directly
  connected to the acceleration: it is mainly produced by accelerated
  electrons, emitting synchrotron radiation. If protons are present in the
  source, they are accelerated in the same way. Thus, the electromagnetic
  energy, $E_{\rm em}$, must correspond to or exceed the total energy observed
  in cosmic rays, $E_{\rm CR}$, $E_{\rm em}\geq E_{\rm CR}$.
Here, the total electromagnetic energy release of a source class can be
determined by multiplying the characteristic luminosity with the life time of
the sources.
To determine the total energy in cosmic rays, the total
energy density measured at Earth needs to be multiplied with the expected
volume of the sources producing the cosmic rays. For a detailed discussion,
see e.g.~\cite{gaisser,julias_review}.
\end{enumerate}
A list
of proposed sources of cosmic rays contributing at different energies in the
spectrum is given in table~\ref{sources_cr_spect:tab}.
%
\begin{table}[h]
\caption{\label{sources_cr_spect:tab}Possible source classes of the cosmic ray spectrum.}
\begin{center}
\begin{tabular}{lllll}
\br
Source class& $L_{\rm em}$ [erg/s]&life time [yr]&energy range&Ref\\ 
\mr
SNR& $10^{42}$&1000&$10^{10}$~eV$<\ep<10^{15}$~eV&\cite{ginzburg1964}\\
SNR-wind&$10^{44}$&1000&$10^{10}$~eV$<\ep<10^{18}$~eV&\cite{voelk_biermann1988}\\
X-ray binaries&$10^{38}$&$10^{5}-10^{6}$&$10^{14}$~eV$<\ep<10^{18}$~eV&\cite{gaisser}\\ 
Pulsars&$10^{37}$&$10^{6}$&$10^{14}$~eV$<\ep<10^{18}$~eV&\cite{gaisser}\\ 
Galaxy clusters&$\sim 10^{44}$&$10^{7}$&$3\cdot 10^{18}$~eV$<\ep< 10^{21}$~eV&\cite{kang1997}\\ 
AGN&$10^{44}-10^{47}$&$10^{7}$&$3\cdot 10^{18}$~eV$<\ep< 10^{21}$~eV&\cite{biermann_strittmatter1987}\\ 
GRBs&$10^{49}-10^{51}$&$10^{-8}-10^{-4}$&$3\cdot 10^{18}$~eV$<\ep< 10^{21}$~eV&\cite{vietri95,wb97}\\
\br
\end{tabular}
\end{center}
\end{table}
Supernova remnants (SNRs) in the Galaxy are the best candidates for cosmic ray
production below the knee of the cosmic ray spectrum, i.e.~$E<10^{15}$~eV. If
a supernova does not explode into the interstellar medium, but into a dense
environment ('SNR-wind'), energies up to $\sim 10^{18}$~eV can be
achieved. X-ray binaries and pulsars in the Galaxy to not provide enough total
energy of the cosmic ray flux below the knee. However, their maximum energy
and their energy output is enough to contribute in the energy range between
knee and ankle, i.e.~$10^{15}$~eV~$<E_p<10^{18}$~eV. Above the ankle, the
sources cannot be of Galactic origin\footnote{At energies of above
$10^{18}$~eV, the energy of the emitted charged particles is high
enough to let the particles travel on straight lines despite 
Galactic magnetic field deflections. Since no such anisotropy
was observed for cosmic rays above the ankle, they have to come from
larger distances, i.e.~they must be extragalactic.}. Extragalactic candidates
providing sufficient maximum energies and total energy release are galaxy
clusters, active galactic nuclei (AGN) and gamma ray bursts (GRBs). 

Alternative neutrino production scenarios are usually connected with {\it
  hidden} cosmic ray sources. If an accelerator of cosmic rays lies behind
dense molecular clouds, the high-energy protons interact with the hydrogen
atoms in the clouds. While the protons are absorbed this way, neutrinos are
produced in high numbers as described above. This process can happen on
Galactic scales, when a supernova remnant or a pulsar sits behind molecular
clouds, or on extragalactic scales, when an AGN jet is absorbed by dense
matter. 
\section{Neutrino flux models\label{nu_models:sec}}
The neutrino flux of a single emitting source follows the protons in
their spectral behavior. The energy spectrum, $d\Phi_{\nu}/dE_{\nu}$, is thus represented by a
power law with a spectral index around $\alpha_{\nu}\sim 2$ and a
cutoff depending on the maximum proton energy, $E_{\nu}^{\max}\sim
E_{p}^{\max}/20$:
\begin{equation}
\frac{d\Phi_{\nu}}{d\en}=A_{\nu}\cdot {\en}^{-\alpha_{\nu}}\cdot \exp\left(-\frac{\en}{{\en}^{\max}}\right)\,.
\end{equation}
The signal strength, $A_{\nu}$, can be determined by assuming
the production of neutrinos in correlation with either high-energy
cosmic rays or photon emission. Depending on the model, neutrinos can
for instance
be connected to radio emission, because it reflects the acceleration
of electrons, which in turn mirrors the protons' acceleration. Alternatively, high-energy photon emission
can be accompanied by a neutrino signal: Neutral pions, originating
from the same process as charged pions as described above, decay into
photons at MeV-TeV energies. 

The neutrino flux from a class of objects, $dN_{\nu}/dE_{\nu}$, can then be determined by
folding the single source flux with the source distribution function,
$dn/dL/dV$, giving the number of sources per luminosity $L$ and per
comoving volume $dV$:
\begin{equation}
\frac{dN_{\nu}}{dE_{\nu}}=\int_z\int_L\,dz\,dL\,\frac{d\Phi_{\nu}}{d\en}\cdot
\frac{dn}{dL\,dV}\cdot \frac{dV}{dz}\cdot \frac{1}{4\pi\,d_{L}^{2}}\,.
\end{equation}
The factor $(4\pi\,d_{L}^{2})^{-1}$ takes into account the
decrease of the flux with the luminosity distance $d_L$. Here, the status of neutrino flux modeling in the context of
observational astronomy is reviewed.
\subsection{The progress of neutrino astronomy so far}
\begin{figure}[h]
\centering{
\includegraphics[width=8.5cm]{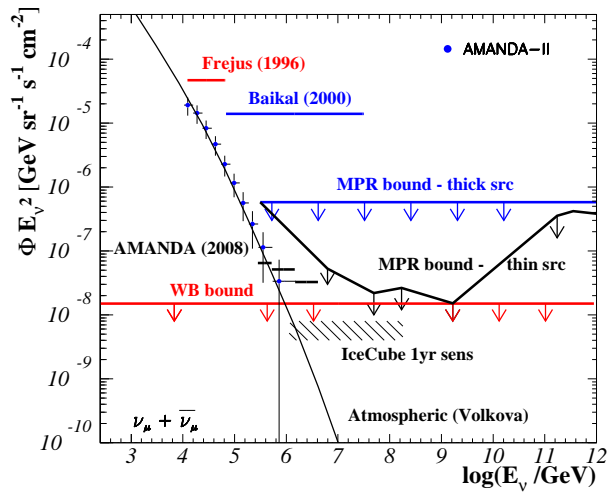}
\caption{High-energy neutrino spectrum. Theoretical bounds:
 upper
  bound on $n\,\gamma-$optically thick (uppermost, blue horizontal line) and
  optically thin (black, broken powerlaw line) sources~\cite{mpr};
  $E^{-2}-$bound on optically thin sources (red, lower horizontal
  line)~\cite{wb97}. Experimental limits: Fr{\'e}jus
  (1996)~\cite{frejus_limit}, Baikal (2000)~\cite{baikal2000}, AMANDA
  (2008) limit and atmospheric spectrum~\cite{kirsten_phd}. The
  atmospheric prediction is given in~\cite{volkova_zatsepin}. IceCube
  sensitivity for 1 full year from~\cite{francis_sept2008}.\label{mpr_wb_procs:fig}}
}
\end{figure}
The modeling of neutrino fluxes constantly improved throughout the years,
using new pieces of information provided by cosmic ray experiments,
keV-TeV photon astronomy and in particular from large volume neutrino
detectors which have improved limits by several orders of magnitude
over the years. Despite the lack of detected sources so far, those
strict experimental limits start to exclude certain classes of
sources, leaving other source classes in favor of producing
high-energy neutrinos. One of the most important examples is that the correlation between
X-rays and neutrinos from AGN can be excluded: the possible contribution of
X-ray emitting AGN to the neutrino background was calculated by
different authors~\cite{nellen,stecker96,alvarez_xrays}. Each of the
predictions exceeds current neutrino flux limits by at least one order
of magnitude, see~\cite{stacking_diff2007}. 

The progress of
  neutrino astronomy within the past 10 years is shown in
  Fig.~\ref{mpr_wb_procs:fig}. Experimental limits were at
  a level $>10^{-5}\,\diffunits$ in the late 1990s and have improved by
  more than 2 orders of magnitude in the past 10 years. This is an
  important step, given the theoretical upper bounds derived in the
  late 1990s as well, see~\cite{wb97,mpr}, implying that sources
  optically thin for neutron-photon interactions cannot provide more
  than $\sim 10^{-6}\,\diffunits$ and optically thick sources are
  restricted to $\lesssim 10^{-6\rightarrow -8}\,\diffunits$, depending on
  the energy. Today, neutrino flux models
  below those two bounds can easily be explored by experiments like
  AMANDA, Anita, Auger and many more. Next
  generation neutrino telescopes like IceCube and KM3NeT will improve
  the sensitivities by a further factor $10$ within the first year of
  operation.
%
\subsection{Prospects for the near future}
\begin{figure}[h]
\centering{
\includegraphics[width=9.cm]{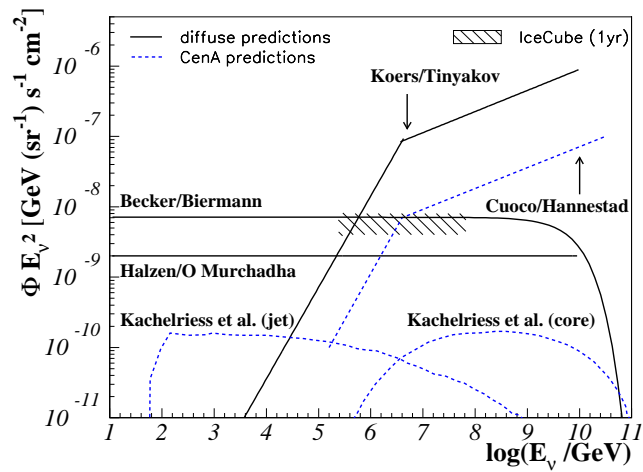}
\caption{Predictions in connection to the hint of anisotropy in
  cosmic rays observed by Auger~\cite{auger_science2007}. Diffuse
  predictions: $p\,\gamma-$ interactions with synchrotron
  photons~\cite{koers_tinyakov2008}; coincident emission of neutrinos and
  TeV $\gamma-$rays~\cite{halzen_omurchadha2008}; neutrinos from the base of
   AGN jets~\cite{becker_biermann2008}. Predictions for CenA:
  $p\,\gamma-$interactions~\cite{cuoco_hannestad2008}; production in the
  AGN jet and core~\cite{kachelriess2008}.\label{auger_neutrinos:fig}}
}
\end{figure}
Many of the source classes are still unexplored due to limited
experimental sensitivities. The two most interesting classes of
extragalactic objects, GRBs and AGN, providing
intensities in excess of IceCube's sensitivity within the first years
of operation, are presented here.
\begin{itemize}
\item {\it Gamma ray bursts}\\
GRBs were proposed as the sources of cosmic rays above the
ankle~\cite{vietri95,wb97}. Due to the high synchrotron field in
those sources, they are good neutrino candidates as
well~\cite{wb97,guetta2004,bshr2006}. The stacking of identified GRBs
for neutrino data provides the best limit so far, see~\cite{kyler_paper},
reaching within a factor of a few above the classical GRB neutrino
flux prediction given in \cite{wb97}. A single, bright GRB can be
identified with a fully operational $1$~km$^3$-detector like
IceCube. The bright burst GRB080319b was observed by IceCube while
under construction, with $\sim 1/10$~km$^3$ operational volume, and $\sim 0.1$
expected events at a much lower background level~\cite{ty_neutrino08}. With a full IceCube detector,
between $1-2$ events are expected.
\item {\it Active galactic nuclei}\\
A variety of AGN sub-classes are proposed as potential neutrino
emitters. Among the most prominent ones are TeV photon emitters, under
the assumption that the TeV emission originates from $\pi^{0}-$decays
rather than from Inverse Compton scattering, see e.g.~\cite{mpr}. Another option is that those AGN responsible for the
cosmic ray flux above the ankle also produce neutrinos. This implies
that the cosmic ray emitters need to provide a target that absorbs
part of the cosmic rays before they can escape the source, i.e.~either
sufficiently high photon or proton fields. Possible targets are the
accretion disk's radiation field, the synchrotron field in the jet, or
the torus as a baryonic target. A first hint of a correlation between
the highest energy cosmic rays and a catalog of close AGN was announced by
Auger~\cite{auger_science2007}. This correlation can be used to
estimate the neutrino flux from the same sources. Different
predictions, many of them based on the assumption that the close AGN
Centaurus A is one of the dominant sources, are presented in Fig.~\ref{auger_neutrinos:fig}. 
\end{itemize}
\section{Outlook\label{outlook:sec}}
Previous high-energy neutrino detectors turned helped excluding
different scenarios of neutrino production so far. Future detectors like IceCube
and KM3NeT will provide
large enough effective volumes to explore further orders of magnitude
in the neutrino energy density of different single sources or source
classes. 

In order to identify the sources of cosmic rays and the mechanisms
producing them, multiwavelength astronomy is crucial. Each of the messengers,
i.e.~charged cosmic rays, radio to TeV photons, neutrinos and also
gravitational waves, provide us with pieces of information. We need to
have all those pieces in order to see the whole picture. The role of
neutrino astronomy in this puzzle is to identify the birth place of cosmic rays.
\section*{Acknowledgments}
Many thanks to my colleagues Francis Halzen, Peter Biermann,
Wolfgang Rhode for inspiring discussions.
Also, I would like to acknowledge the support from the grant
BMBF~05CI5PE1/0.
\providecommand{\newblock}{}


\begin{thebibliography}{10}
\expandafter\ifx\csname url\endcsname\relax
  \def\url#1{{\tt #1}}\fi
\expandafter\ifx\csname urlprefix\endcsname\relax\def\urlprefix{URL }\fi
\providecommand{\eprint}[2][]{\url{#2}}

\bibitem{hess1912}
{Hess} V~F 1912 {\em Phys.~Z.\/} {\bf 13} 1084

\bibitem{julias_review}
{Becker} J~K 2008 {\em Physics Reports\/} {\bf 458} 173 (\textit{Preprint}
  \eprint{ArXiv:0710.1557})

\bibitem{hillas1984}
{Hillas} A~M 1984 {\em Ann.~Rev.~Astron.~Astrophys.\/} {\bf 22} 425

\bibitem{gaisser}
Gaisser T~K 1990 {\em Cosmic Rays and Particle Physics\/} (Cambridge University
  Press)

\bibitem{ginzburg1964}
{Ginzburg} V~L and {Syrovatskii} S~I 1964 {\em {The Origin of Cosmic Rays}\/}
  (Pergamon Press)

\bibitem{voelk_biermann1988}
{V\" olk} H~J and {Biermann} P~L 1988 {\em Astroph.~Journal Let.\/} {\bf 333}
  L65

\bibitem{kang1997}
{Kang} H, {Rachen} J and {Biermann} P~L 1997 {\em Mon.~Not.~of the Royal
  Astron. Soc.\/} {\bf 286} 257

\bibitem{biermann_strittmatter1987}
{Biermann} P~L and {Strittmatter} P~A 1987 {\em Astroph.~Journal\/} {\bf 322}
  643

\bibitem{vietri95}
{Vietri} M 1995 {\em Astroph.~Journal\/} {\bf 453} 883 (\textit{Preprint}
  \eprint{astro-ph/9506081})

\bibitem{wb97}
{Waxman} E and {Bahcall} J~N 1997 {\em Phys.~Rev.~Let.\/} {\bf 78} 2292

\bibitem{mpr}
{Mannheim} K, {Protheroe} R~J and {Rachen} J~P 2001 {\em Phys.~Rev.~D\/} {\bf
  63} 23003

\bibitem{frejus_limit}
{Rhode} W and {Daum} K 1996 {\em Astropart.~Phys.\/} {\bf 4} 217

\bibitem{baikal2000}
{Balkanov} V~A, {(Baikal Coll)} {\em et~al.\/} 2000 {\em Astropart.~Phys.\/}
  {\bf 14} 61

\bibitem{kirsten_phd}
M{\"u}nich K 2007 Ph.D. thesis Universit\"at Dortmund

\bibitem{volkova_zatsepin}
{Volkova} L~V and {Zatsepin} G~T 1983 {\em Soviet J.~of Nuclear Physics\/} {\bf
  37} 212

\bibitem{francis_sept2008}
{Halzen} F 2008  (\textit{Preprint} \eprint{ArXiv:0809.1874})

\bibitem{nellen}
{Nellen} L, {Mannheim} K and {Biermann} P~L 1993 {\em Phys.~Rev.~D\/} {\bf 47}
  5270

\bibitem{stecker96}
{Stecker} F~W and {Salamon} M~H 1996 {\em Space Science Rev.\/} {\bf 75} 341

\bibitem{alvarez_xrays}
{Alvarez-Mu{\~n}iz} J and {M{\'e}sz{\'a}ros} P 2004 {\em Phys.~Rev.~D\/} {\bf
  70} 123001 (\textit{Preprint} \eprint{astro-ph/0409034})

\bibitem{stacking_diff2007}
{Becker} J~K {\em et~al.\/} 2007 {\em Astropart.~Phys.\/} {\bf 28} 98
  (\textit{Preprint} \eprint{arXiv:astro-ph/0607427})

\bibitem{auger_science2007}
{Auger Collaboration} 2007 {\em Science\/} {\bf 318} 938 see also
  www.sciencemag.org

\bibitem{koers_tinyakov2008}
{Koers} H~B~J and {Tinyakov} P 2008  (\textit{Preprint}
  \eprint{ArXiv:0802.2403})

\bibitem{halzen_omurchadha2008}
{Halzen} F and {O'Murchadha} A 2008  (\textit{Preprint}
  \eprint{ArXiv:0802.0887})

\bibitem{becker_biermann2008}
{Becker} J~K and {Biermann} P~L 2008 {\em submitted to Astrop.~Phys.\/}
  (\textit{Preprint} \eprint{ArXiv:0805.1498})

\bibitem{cuoco_hannestad2008}
{Cuoco} A and {Hannestad} S 2008 {\em Phys.~Rev.~D\/} {\bf 78} 023007
  (\textit{Preprint} \eprint{arXiv:0712.1830})

\bibitem{kachelriess2008}
{Kachelriess} M, {Ostapchenko} S and {Tomas} R 2008  (\textit{Preprint}
  \eprint{0805.2608})

\bibitem{guetta2004}
{Guetta} D {\em et~al.\/} 2004 {\em Astropart.~Phys.\/} {\bf 20} 429

\bibitem{bshr2006}
{Becker} J~K {\em et~al.\/} 2006 {\em Astropart.~Phys.\/} {\bf 25} 118

\bibitem{kyler_paper}
{Achterberg} A, {(IceCube Coll)} {\em et~al.\/} 2008 {\em Astroph.~Journal\/}
  {\bf 674} 357

\bibitem{ty_neutrino08}
{DeYoung} T, {(IceCube Coll)} {\em et~al.\/} 2008 These proceedings

\end{thebibliography}
\end{document}